\documentclass[twocolumn, floatfix, superscriptaddress,dvipdfmx]{revtex4-1}
\usepackage{amsmath,bm,graphicx}
\usepackage{amsmath}
\usepackage{amssymb}
\usepackage{amsfonts}
\begin{document}
\title{Bouncing of a projectile impacting a dense potato-starch suspension layer}

\author{Kazuya Egawa}
\author{Hiroaki Katsuragi}
\affiliation{Department of Earth and Environmental Sciences, Nagoya University, Nagoya 464-8601, Japan}

\date{\today}

\begin{abstract}
When a solid projectile is dropped onto a dense non-Brownian-particle suspension, the action of an extremely large resistance force on the projectile results in its drastic deceleration, followed by a rebound. In this study, we perform a set of simple experiments of dropping a solid-projectile impact onto a dense potato-starch suspension. From the kinematic data of the projectile motion, the restitution coefficient and timescale of the rebound are measured. By assuming linear viscoelasticity, the effective transient elasticity and viscosity can be estimated. We additionally estimate the Stokes viscosity on a longer timescale by measuring the slow sinking time of the projectile. The estimated elastic modulus and viscosity are consistent with separately measured previous results. In addition, the effect of mechanical vibration on the viscoelasticity is examined. As a result, we find that the viscoelasticity of the impacted dense suspension is not significantly affected by the mechanical vibration. 
\end{abstract}
\maketitle

\section{Introduction}
\label{sec:introduction}
Running on fluid is not impossible if the fluid is so-called dilatant fluid such as a dense mixture of water and corn starch or potato starch. This type of dense suspension shows a sudden increase in viscosity at a certain shear strain rate~\cite{Fall:2008,Madraki:2018}. This discontinuous shear thickening (DST) has been considered the possible reason for the effective hardening of the dense suspension. Frictional interaction between grains plays an essential role in producing DST~\cite{Seto:2013}. Shear thickening in a dense suspension has attracted the interest of many physicists~\cite{Brown:2014}. In particular, DST is a very intriguing phenomenon observed in a sheared dense suspension. 
However, DST has been mainly measured by steady-state rheometry, although running on dense suspension induces a highly transient response against the impact of the foot. To directly mimic the running situation, Waitukaitis and Jaeger performed impact experiments and revealed that the effectively solidified region is developed in the impacted suspension by dynamic jamming-front propagation~\cite{Waitukaitis:2012}. They proposed the added-mass effect owing to this effectively solidified zone, and concluded that the added mass results in a large deceleration of the impactor. Then, dynamic jamming-front propagation was directly observed in a dry granular system~\cite{Waitukaitis:2013} of two-dimensional (2D)~\cite{Peters:2014} and three-dimensional (3D)~\cite{Han:2016} dense suspensions. The solid plug made by the impact was indirectly observed by the solid indentation~\cite{Liu:2010}. Although these results indicate the importance of dynamic jamming-front propagation in impacted dense suspension, the induced added-mass effect is actually insufficient to support a person running on dense suspension~\cite{Mukhopadhyay:2018}. 

To overcome this difficulty, the importance of elasticity and the role of the boundary have been discussed recently on the basis of constant-rate penetration experiments~\cite{Allen:2018,Maharjan:2018}. According to their idea, when the dynamically jammed region reaches the boundary of the container, it causes an effective elasticity. However, when the dynamic jamming front propagates in an infinite space, the solidified region can only contribute to an increase in added mass. In an impact situation, while the effect of added mass can decelerate the projectile, it never causes a rebound by the effective elasticity. Note that, in this paper, we use the term ``elasticity'' to merely express the solid-like property, which does not refer to energy storage.  
In addition, even brittle fracturing has been observed on impacted very shallow dense suspension~\cite{Smith:2010,Roche:2013,Allen:2018}. These experimental results suggest the significance of solid-like behavior in an impacted dense suspension. 

The complex rheological properties of a dense suspension result in various interesting phenomena, e.g., the stop-and-go (oscillatory) motion of a sinking ball in dense suspension~\cite{vonKann:2011,vonKann:2013} and oscillation in a rotating system~\cite{Nagahiro:2016}. These oscillatory behaviors may originate from the hysteresis of viscosity in a dense suspension~\cite{Deegan:2010} or the capillary effect~\cite{Jerome:2016}. Moreover, mechanical vibration also affects the rheological properties of dense suspension. Both liquefaction~\cite{Hanotin:2012,Clement:2018} and solidification (proved by the stable holes) can be induced by vibration~\cite{Merkt:2004}. That is, the mechanical properties of a dense suspension can vary depending on perturbations such as mechanical vibration. 

To quantitatively characterize the elasticity of dense suspension, an analysis of the rebound using a solid-projectile impact could be helpful. Indeed, the rebound of a projectile was observed in some previous studies on dense suspension~\cite{Waitukaitis:2012,Waitukaitis:2013PhD,Peters:2016}. However, a detailed analysis of a rebound has not yet been performed. 

The study of impact drag is one of the most useful methods to characterize the transient rheology of various soft matters~\cite{Katsuragi:2016}.  
For the rheological characterization of dense suspension, (steady-state) viscosity has long been measured. However, we are interested not only in the viscosity but also the elasticity. Thus, in this study, we employ the simplest viscoelastic (Voigt) model to analyze the transient drag-force rheology of dense suspension. Namely, we develop a simple method to estimate the transient viscoelasticity of the impacted dense suspension as based on a simple linear model. In addition, by measuring the slow sinking timescale of the projectile, we estimate the viscosity of the dense suspension on a longer timescale that should be significantly different from the transient one. Moreover, the effect of vibration on the viscoelastic behavior of the impacted dense suspension is also investigated. As already mentioned, mechanical vibration can soften or harden the dense suspension. To evaluate the vibrational effect, we simply apply mechanical vibration to the target dense suspension and measure its viscoelasticity through rebound analysis.

\section{Experiment}
In this study, we conduct a set of experiments of a solid-projectile impact onto a surface of a dense-suspension target and precisely measure the response of the projectile. Specifically, we measure the restitution coefficient, rebound timescale, and slow sinking timescale by using the kinematic data of the projectile.

\begin{figure}
	\centering
	\includegraphics[width=1.\linewidth]{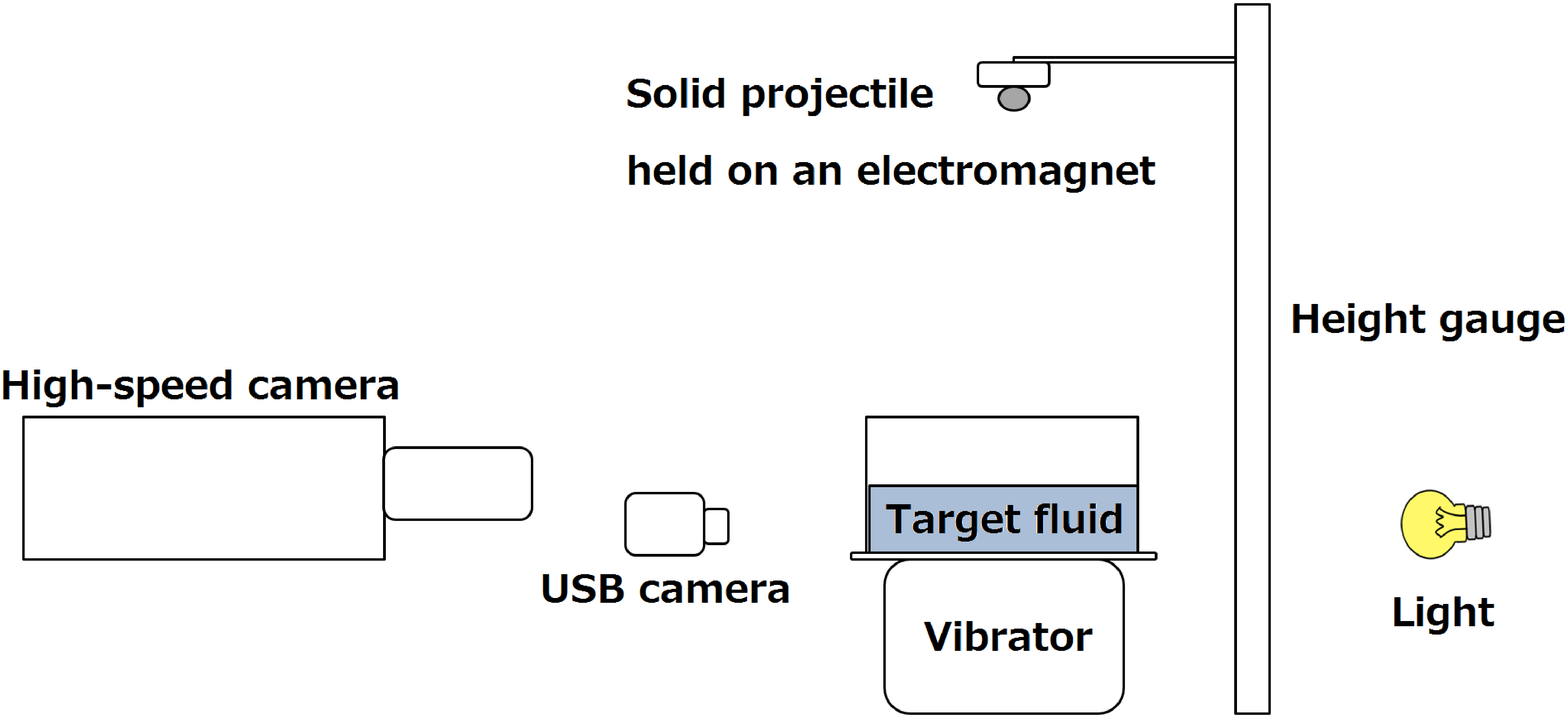}
	\caption{Schematic image of experimental setup. An iron steel sphere is released from a certain height onto a target of dense potato-starch suspension. Motion of projectile is captured by high-speed camera. Slow sinking timescale is also measured by USB camera. To evaluate effect of vibration, target container is mounted on a vibrator.}
	\label{fig:Experiment}
\end{figure}

The experimental setup is illustrated in Fig.~\ref{fig:Experiment}. An aqueous dense suspension of potato starch (Maruboshi, true density $\rho_{\rm s} \simeq  1.41~\times~10^3~\rm{kg~m^{-3}}$) is prepared by mixing it with purified water. The mass ratio of potato starch to purified water is fixed at 1.5 (packing fraction $\phi \simeq 0.51$) in this study. For example, 100~g of purified water and 150~g of potato starch are mixed with a spoon. Then, the suspension is poured into a square acrylic container (inner dimensions: 100 mm $\times$ 100 mm $\times$ 60 mm). The thickness of the target layer is varied in the range of $10~{\rm mm} \leq H \leq 40~{\rm mm}$. In each experiment, an iron steel sphere (diameter $D_{p} = 8$~mm, density $\rho_{p} =  8 \times 10^3~\rm{kg~m^{-3}}$) is released from a certain height by using an electromagnet. The released (free-fall) height range is $10~\rm{mm} \leq$ $h_{\rm imp}$ $\leq 150~{\rm mm}$. The released projectile impinges on the surface of the suspension. The corresponding impact velocity range is $0.4~\rm{m~s^{-1}} \leq$ $v_0$ $\leq 1.7~\rm{m~s^{-1}}$. Before each impact, we manually stirred the suspension to erase the memory of the prior impact and to prepare a homogeneous suspension. The projectile is released right after stirring the target in order to minimize the effect of precipitation. Since the impact occurs in a very short duration (less than 1~s), we neglect the effect of precipitation in this experiment. A high-speed camera (Photron SA-5) is used to capture the motion of the projectile at 12,000 frames per second ($896 \times 848~{\rm pixels}$. The spatial resolution is $25~{\rm \mu m} / {\rm pixel}$). We also use a USB camera (STC-MCCM401U3V) to film the slow sinking of the projectile on a longer timescale, at 50 frames per second. 
To examine the vibration effect, the target-fluid container is mounted on a vertical vibrator (Daiei, angel vibrator digital or Emic corporation, vibration generator, 513-B/A). The maximum vibration acceleration $a_{\rm vib}$ measured by an accelerometer (Emic corporation, vibration pickup, 710-D) is varied in the range of $0~{\rm m~s^{-2}} \leq a_{\rm vib} \leq 200~{\rm m~s^{-2}}$. A sinusoidal vertical vibration $a_{\rm v}(t)=a_{\rm vib}{\sin}(2\pi ft)$ is applied to the container. In this vibration experiment, the frequency $f$ is fixed at $f~=~120~{\rm Hz}$. The thickness of the target fluid is also fixed at $H=20$~mm in the vibration experiment. We carry out at least three experimental runs for each experimental condition to check the reproducibility. Error bars in the plots of this paper represent the standard deviation of the multiple experimental runs. The typical order of the rebounding distance is $O(10^{-1})$~mm, and the order of measurement resolution is $O(10^{-2})$~mm. Thus, the measurement uncertainty is several tens of percent. This uncertainty is smaller than the variation of target conditions among repeated experiments. Thus, the error is dominated by the standard deviation among the repeated experimental runs. 

\section{Results and analyses}
First, we present the typical raw data and the representative result by focusing on the impacts without target vibration. Then, the result of the vibration effect is presented in Sec.~\ref{sec:vibration} after explaining all of the analysis methods.

\subsection{Raw data}
\begin{figure*}
	\centering
	\includegraphics[width=1.\linewidth]{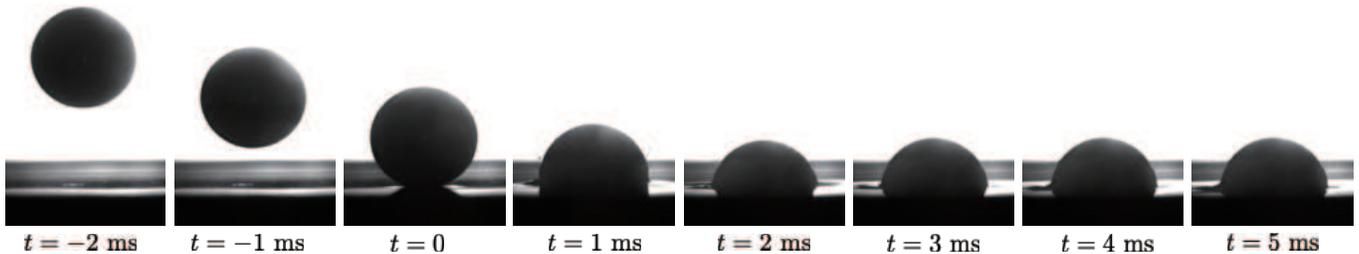}
	\caption[Raw data images]{
Successive raw data images of projectile motion at every 1~ms for $D_{p}=8$~mm, $H = 20~{\rm mm}$, $v_0 = 1.35~{\rm m~s^{-1}}$, and without target vibration.}
	\label{fig:Raw-data-images}
\end{figure*}

\begin{figure}
	\centering
	\includegraphics[clip,width=.9\linewidth]{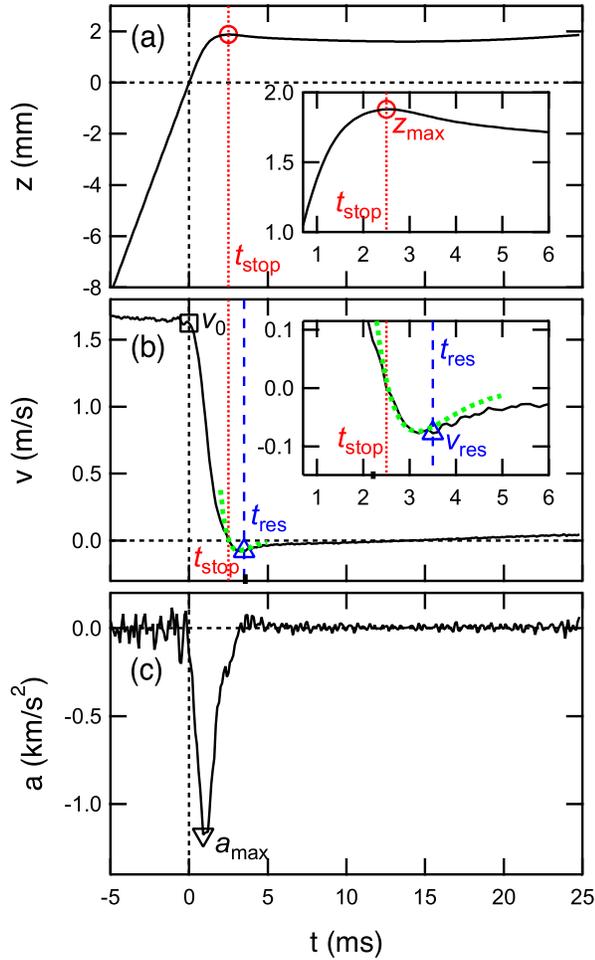} 
	\caption{Example data of projectile motion around impact moment (vertically downward direction is positive). Experimental conditions are $D_{p}=8$~mm, $H = 20~{\rm mm}$, $v_0 = 1.62~{\rm m~s^{-1}}$, and without target vibration. (a)~Position $z(t)$, (b)~velocity $v(t)$, and (c)~acceleration $a(t)$ are presented. Inset of panel (a) shows magnified $z(t)$ around maximum penetration depth $z_{\rm max}$ (circular symbol). Inset of panel (b) shows magnified $v(t)$ around restitution velocity $v_{\rm res}$ (triangular symbol). Square and inverted-triangle symbols in panels (b) and (c) indicate impact velocity $v_0$ and maximum absolute acceleration $a_{\rm max}$, respectively. Timescales $t_{\rm stop}$ (defined by $v=0$) and $t_{\rm res}$ (defined by $v=v_{\rm res}$) are also shown in panel (b). Fitting to model of Eq.~(\ref{eq:VoigtModel}) is displayed as green dotted curve in panel (b).}
	\label{fig:Kinematic-data-sets}
\end{figure}

Example images of projectile motion acquired by the high-speed camera are shown in Fig.~\ref{fig:Raw-data-images}. As a representative case, a $H=20$~mm target without vibration is shown in Fig.~\ref{fig:Raw-data-images} (and Fig.~\ref{fig:Kinematic-data-sets}). The interval between the successive images is 1~ms in Fig.~\ref{fig:Raw-data-images}. One can confirm that the projectile penetrates and slightly rebounds just after the maximum penetration. 

\begin{figure}
	\centering
	\includegraphics[width=.9\linewidth]{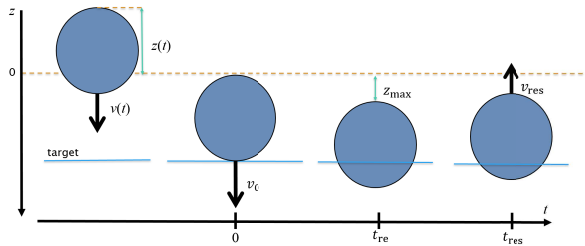}
	\caption{Schematic diagram of image analysis method with definitions of characteristic quantities. Top position of projectile is measured to analyze motion.}
	\label{fig:method}
\end{figure}

	To examine the motion of the projectile, we conduct an image analysis. Figure~\ref{fig:method} schematically illustrates the image analysis method. We identify the moment of impact at which the projectile comes into contact with the target. The reference time $t=0$ is defined by this impact moment. The position of the top of the projectile is denoted as $z$ ($z=0$ at $t=0$, and the vertically downward direction is defined as the positive direction). Then, the velocity $v(t)$ and acceleration $a(t)$ are calculated from the temporal differentiations of $z$: $v(t) = dz(t)/dt$ and $a(t) = dv(t)/dt$. From the kinematic data, we measure the following characteristic quantities: maximum penetration depth $z_{\rm max}$, impact velocity $v_0$, stopping time $t_{\rm stop}$, restitution time $t_{\rm res}$, restitution velocity $v_{\rm res}$, and peak deceleration magnitude $a_{\rm max}$. Using these quantities, we will analyze the dynamics of the projectile and corresponding rheological properties of the target dense suspension.

Example data of the projectile motion derived from the image analysis are shown in Fig.~\ref{fig:Kinematic-data-sets}. The position $z(t)$ (Fig.~\ref{fig:Kinematic-data-sets}(a)), velocity $v(t)$ (Fig.~\ref{fig:Kinematic-data-sets}(b)), and acceleration $a(t)$ (Fig.~\ref{fig:Kinematic-data-sets}(c)) are shown as functions of time $t$. We define the maximum penetration depth $z_{\rm max}$ by the maximum value of $z(t)$ right after the impact (inset of Fig.~\ref{fig:Kinematic-data-sets}(a)). Likewise, we define $v_{\rm res}={\rm max}(|v|)$ and $a_{\rm max}={\rm max}(|a|)$ by the absolute maximum values of $v$ and $a$ after the impact, respectively (Fig.~\ref{fig:Kinematic-data-sets}(b,c)). $t_{\rm stop}$ is defined by the moment of $v=0$ after the impact. $t_{\rm res}$ corresponds to the time of $v=v_{\rm res}$. In Fig.~\ref{fig:Kinematic-data-sets}, we observe a clear rebound after the impact. Basically, similar bouncing behaviors are observed in most of the impacts in this study. However, impacts without rebounds can be observed for a thick target ($H = 40$~mm) with a low impact velocity ($v_0<0.8$~m~s$^{-1}$). In this study, we analyze the rebound data because we are interested in the viscoelastic behavior. The rebound is necessary to characterize the elasticity, as discussed in Sec.~\ref{sec:VoigtModel}.

\subsection{Maximum penetration depth and maximum acceleration}
\label{sec:max-z-max-a}
\begin{figure}
	\centering
	\includegraphics[width=.9\linewidth]{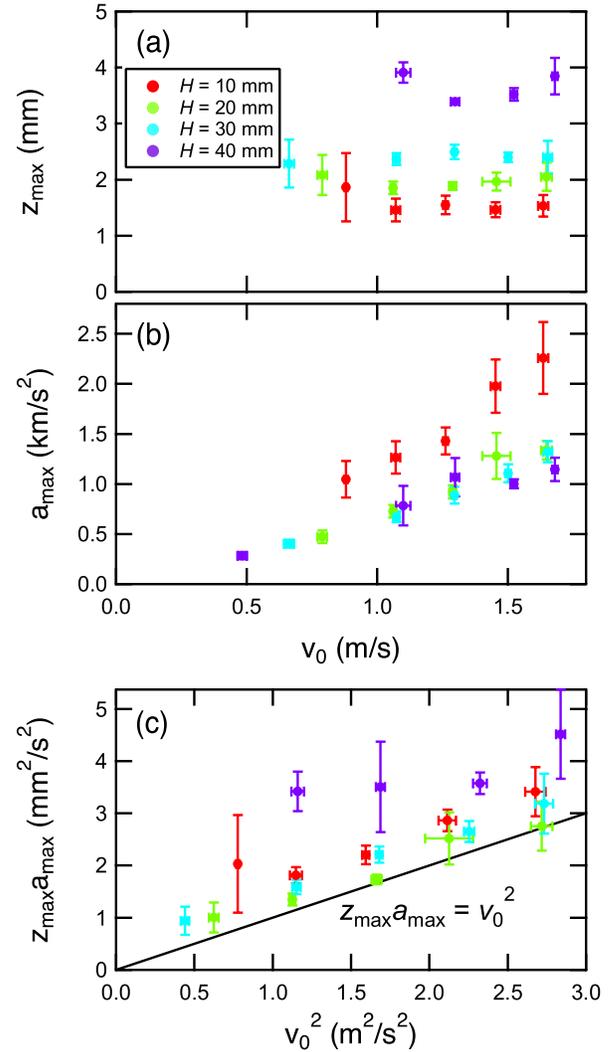} 
	\caption[Maximum penetration depth and maximum acceleration]
	{(a)~Maximum penetration depth $z_{\rm max}$ vs. impact velocity $v_0$ and (b)~maximum acceleration $a_{\rm max}$ vs. $v_0$. Color code indicates variation of thickness of target layer $H$ as denoted in legend. Same color code is used in all plots. (c)~Relation between $z_{\rm max}a_{\rm max}$ and $v_{0}^2$. Straight line indicates linear relation $z_{\rm max}a_{\rm max} = v_0^2$.}
	\label{fig:zmax-amax}
\end{figure}

Relations between $z_{\rm max}$, $a_{\rm max}$, and $v_0$ are shown in Fig.~\ref{fig:zmax-amax}(a,b). The color code in Fig.~\ref{fig:zmax-amax} is used to indicate $H$. An identical color code is used in all other plots in this paper. As can be seen in Fig.~\ref{fig:zmax-amax}(a), $z_{\rm max}$ is almost independent of $v_0$ and increases with $H$. As shown in Fig.~\ref{fig:zmax-amax}(b), $a_{\rm max}$ increases with $v_0$. 
Figure~\ref{fig:zmax-amax}(c) shows the relation between $z_{\rm max} a_{\rm max}$ and $v_{0}^{2}$. The straight line in Fig.~\ref{fig:zmax-amax}(c) indicates a simple kinematic relation $z_{\rm max} a_{\rm max} = v_0^{2}$ that has been confirmed in a shallow impact onto a dust aggregate~\cite{Katsuragi:2017}. From simple energy conservation, a relation 
\begin{equation}
z_{\rm max}a_{\rm max}=\frac{v_{0}^{2}}{2}
\label{eq:zav}
\end{equation}
can be derived when we assume the constant $a_{\rm max}$ during the deceleration~\cite{Katsuragi:2018}. However, Fig.~\ref{fig:zmax-amax} shows a relation $z_{\rm max}a_{\rm max} \geq v_{0}^2$. 

\subsection{Restitution coefficient and rebound timescale}
\begin{figure}
	\centering
	\includegraphics[width=1.\linewidth]{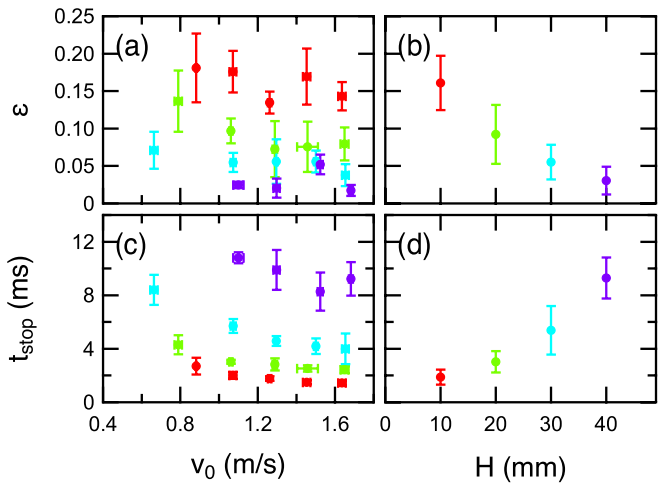} 
	\caption[Restitution coefficient and rebound timescale]{
	Restitution coefficient $\varepsilon$ and stopping timescale $t_{\rm stop}$ measured from kinematic data: (a)~relation between $\varepsilon$ and impact velocity $v_0$, (b)~relation between $\varepsilon$ and thickness of target layer $H$, (c)~relation between $t_{\rm stop}$ and $v_0$, and (d)~relation between $t_{\rm stop}$ and $H$. To compare $H$ dependence, all $v_0$ data (with identical $H$) are averaged in panels (b) and (d). Same average is also taken in following plots with $H$ abscissa.}
	\label{fig:Restitution-coefficient-and-rebound-timescale}
\end{figure}

To further quantify the kinematic data, we measure the restitution coefficient and rebound timescale. From the velocity data (Fig.~\ref{fig:Kinematic-data-sets}(b)), we compute the restitution coefficient $\varepsilon$ as
\begin{equation}
	\varepsilon = \left| \frac{v_{\rm res}}{v_{\rm 0}} \right|.
	\label{eq:restitution-coefficient}
\end{equation}
	In addition, we regard $t_{\rm stop}$ as a characteristic timescale of the deceleration. The measured relations among $\varepsilon$, $t_{\rm stop}$, $v_0$, and $H$ are shown in Fig.~\ref{fig:Restitution-coefficient-and-rebound-timescale}. The top row exhibits $\varepsilon$ and the bottom row exhibits $t_{\rm stop}$ data. The left and right columns respectively show the $v_0$ and $H$ dependences of these quantities. Figure~\ref{fig:Restitution-coefficient-and-rebound-timescale}(a) shows that $\varepsilon$ is almost independent of $v_0$. On the other hand, $t_{\rm stop}$ has a slightly negative correlation with $v_0$ (Fig.~\ref{fig:Restitution-coefficient-and-rebound-timescale}(c)). $H$ dependences of $\varepsilon$ and $t_{\rm stop}$ can be clearly observed. $\varepsilon$ decreases with an increase in $H$ (Fig.~\ref{fig:Restitution-coefficient-and-rebound-timescale}(b)). By contrast, $t_{\rm stop}$ increases with an increase in $H$. Therefore, the thicker the target layer, the later the projectile begins to rebound. In addition, a thicker target layer is more dissipative than a thinner one.

\subsection{Linear dissipative rebound model}
\label{sec:VoigtModel}
Since most of the impacts show small but finite rebounds of the projectile, the effective elasticity should be considered for the constitutive relation of the impacted dense suspension. Here, we consider a simple linear viscoelastic (Voigt) model to characterize the elasticity as well as the viscosity. The equation of the motion of the projectile is written as
\begin{equation}
m_p\frac{d^2z}{dt} = -k_{\rm D}z -\eta_{D}D_p\frac{dz}{dt},
\label{eq:VoigtModel}
\end{equation}
where $m_p$, $k_D$, and $\eta_D$ are the mass of the projectile, effective spring constant of suspension, and effective viscosity of suspension, respectively. In this model, we neglect the effect of gravity (and buoyancy) since the level of deceleration $a_{\rm amx}\simeq 10^3$~m~s$^{-2}$ (Fig.~\ref{fig:Kinematic-data-sets}(c)) is much greater than the earth's gravity. In addition, the effect of the squeeze flow and capillary force are neglected. Because we consider the effective solidification, we simply assume that the squeeze flow is negligible, and the capillary effect can be negligible when the projectile is not very small~\cite{Gollwitzer:2012}. However, the squeeze flow effect is dominant in a narrow-gap situation~\cite{VazquezQuesada:2018}. In a very shallow case (small $H$), the effect might play a certain role. Furthermore, the capillary effect is crucial for the late-stage motion. In this sense, the model of Eq.~(\ref{eq:VoigtModel}) characterizes the ``effective elasticity'' and ``effective viscosity'' causing a dissipative rebound. Assuming a cylindrical-pillar spring with diameter $D_p$, the elastic modulus can be approximately estimated as $E_D=4Hk_D/\pi D_p^2$~(Hooke's law). Using this relation, we can compute the elasticity of the impacted dense suspension once $k_D$ is estimated from the experimental data. 

\begin{figure}
	\centering
	\includegraphics[width=1.\linewidth]{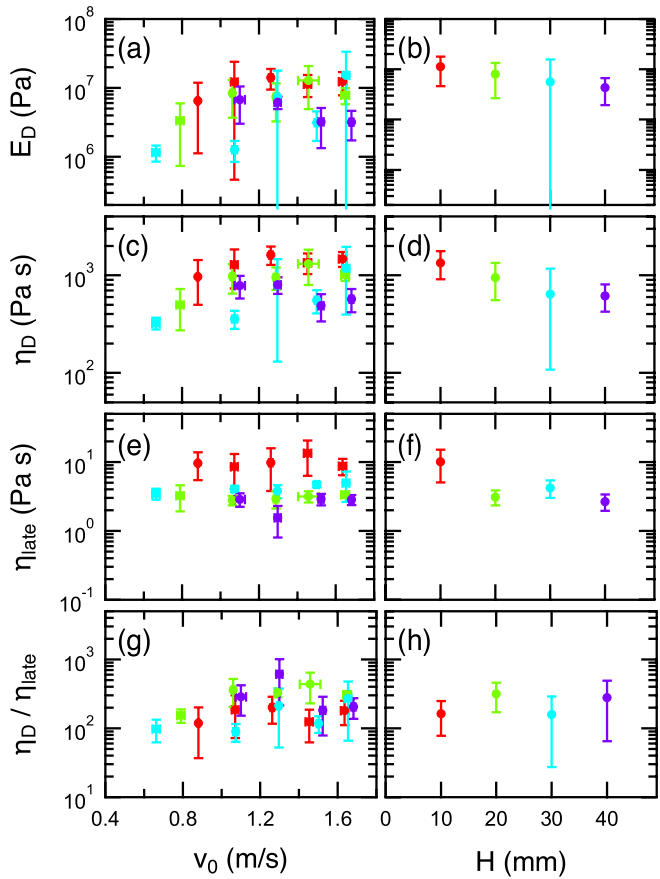} 
	\caption{
$E_D$, $\eta_D$, $\eta_{\rm late}$, and $\eta_D/\eta_{\rm late}$ computed by Eqs.~(\ref{eq:E-ans}), (\ref{eq:eta-ans}), and Stokes drag law. (a)~$E_D$ vs. $v_0$, (b)~$E_D$ vs. $H$, (c)~$\eta_D$ vs. $v_0$, (d)~$\eta_D$ vs. $H$, (e)~$\eta_{\rm late}$ vs. $v_0$, and (f)~$\eta_{\rm late}$ vs. $H$ are shown in each corresponding panel. In panels (g) and (h), viscosity ratio between short timescale and long timescale (above and below DST transition), $\eta_D/\eta_{\rm late}$, is shown.
}
	\label{fig:viscoelastic-results}
\end{figure}

Equation~(\ref{eq:VoigtModel}) is the simplest linear viscoelastic model that can characterize the dissipative rebound. By assuming that the rebound occurs at the half cycle of the attenuating oscillation (in the solution of Eq.~(\ref{eq:VoigtModel})) under the initial (impact) conditions $v=v_0$ and $z=0$ at $t=0$, the restitution coefficient $\varepsilon$ can be written as
\begin{equation}
\varepsilon = \exp\left(-\frac{\pi \eta_{D} D_{p}}{\sqrt{4m_{p}k_{D}-\eta_{D}^{2}D_{p}^{2}}}\right).
\label{eq:epsilon-model}
\end{equation}
The restitution coefficient in this linear model is independent of $v_0$. The corresponding rebound timescale (period of attenuating oscillation) $T$ is derived as
\begin{equation}
T = 2\pi\sqrt{\frac{4m_{p}^{2}}{4m_{p}k_{D}-\eta_{D}^{2}D_{p}^2}}.
\label{eq:T-model}
\end{equation}
These two characteristic quantities can be measured from the experimental data. Then, the effective elastic modulus $E_D=4Hk_D/\pi D_p^2$ and effective viscosity $\eta_D$ are obtained as
\begin{equation}
E_D=\frac{16 \pi m_p H}{D_p^2T^2}\left[1+\left(\frac{\ln \varepsilon}{\pi}\right)^2 \right],
\label{eq:E-ans}
\end{equation}
\begin{equation}
\eta_D = \frac{4m_p}{D_pT}|\ln \varepsilon |.
\label{eq:eta-ans}
\end{equation}
Actually, it is not easy to determine the rebound timescale $T$ from the kinematic dataset. Here, we simply assume that the duration between $t_{\rm stop}$ (time of $v=0$) and $t_{\rm res}$ (time of $v=v_{\rm res}$) corresponds to a quarter of an oscillation period: $T/4=t_{\rm res}-t_{\rm stop}$ (Fig.~\ref{fig:Kinematic-data-sets}(b)). Then, substituting the experimentally obtained $\varepsilon$ and $T$ (and known experimental conditions $D_p$, $m_p$, and $H$) into Eqs.~(\ref{eq:E-ans}) and (\ref{eq:eta-ans}), $E_D$ and $\eta_D$ are computed as shown in Fig.~\ref{fig:viscoelastic-results}(a-d). As can be seen in Fig.~\ref{fig:viscoelastic-results}(a,c), $E_D$ and $\eta_D$ are almost independent of $v_0$ (at least when $v_0>1.0$~m s$^{-1}$). The $v_0$-independent $\varepsilon$ is consistent with the model~(Eq.~(\ref{eq:epsilon-model})). Although $E_D$ and $\eta_{D}$ show a slightly decreasing trend against $H$ (Fig.~\ref{fig:viscoelastic-results}(b,d)), the variations are not very significant. In Fig.~\ref{fig:Kinematic-data-sets}(b), the fitting curve by the model of Eq.~(\ref{eq:VoigtModel}) (with the estimated $E_D$ and $\eta_D$) is shown as a green dotted curve. Although the duration from $t_{\rm stop}$ to $t_{\rm res}$ can be well fitted, the fitting deviates from the data before $t_{\rm stop}$ and after $t_{\rm res}$.

\subsection{Slow sinking timescale}
\begin{figure}
	\centering
	\includegraphics[clip,width=1.\linewidth]{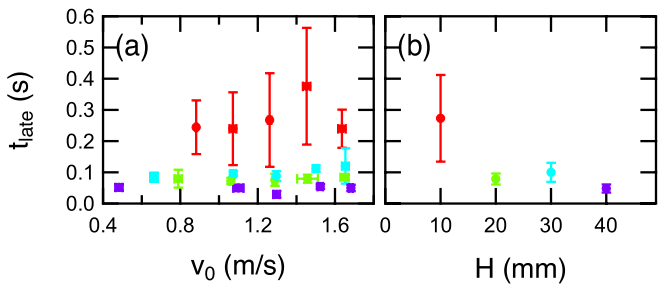} 
	\caption[Slow sinking timescale]{
Slow sinking timescale measured by USB camera. (a)~Relation between slow sinking timescale $t_{\rm late}$ and impact velocity $v_0$. (b)~Relation between $t_{\rm late}$ and thickness of target fluid $H$.}
	\label{fig:Slow-sinking-timescale}
\end{figure}

Thus far, we have used the short timescale data ($\sim10^{-3}~{\rm s}$) to characterize the immediate response of the impacted suspension (above the DST transition) since a sudden stop followed by a rebound has basically been observed. On a longer timescale ($\sim10^{-1}~{\rm s}$), on the other hand, the projectile begins to penetrate again at a relatively slow penetration speed. In this late stage, the dense suspension can be regarded as a typical viscous suspension. To characterize these relatively slow dynamics (below the DST transition), we measure the slow sinking timescale $t_{\rm late}$ from a movie obtained by the USB camera. Since the sinking timescale is too long to be measured by the high-speed camera, we cannot track the motion of the projectile. Thus, we only measure the sinking timescale by using low-resolution data. Here, $t_{\rm late}$ is defined by the moment at which the entire projectile is submerged in the suspension. 
The relations between $t_{\rm late}$ and $v_0$ or $H$ are shown in Fig.~\ref{fig:Slow-sinking-timescale}. As confirmed in Fig.~\ref{fig:Slow-sinking-timescale}, $t_{\rm late}$ is almost independent of $v_0$ and $H$ (at least in the regime $H \geq 20~{\rm mm}$). For the thinner layer ($H = 10~{\rm mm}$), values of $t_{\rm late}$ are slightly greater than in the other cases. 
This slowdown of the penetration could come from the too-close bottom-boundary effect (or squeeze flow effect) rather than the intrinsic viscosity since $D_{p}=8$~mm is close to $H=10$~mm. The order of the typical sinking timescale, $10^{-1}$~s, is much longer than the rebound timescale $10^{-3}$~s (Fig.~\ref{fig:Restitution-coefficient-and-rebound-timescale}). We do not observe any oscillatory motion of the slowly sinking projectile probably owing to the limited target thickness $H$. 

For an analysis of late-stage sinking, we use a viscous terminal velocity on the basis of Stokes' drag law, $\eta_{\rm late}=(\rho_{p}-\rho_{f})gD_{p}^2/18{v_{\eta}}$, where $\rho_f$ is the density of the target fluid. Then, the sinking timescale $t_{\rm late}$ and the maximum penetration depth $z_{\rm max}$ can be transformed into $\eta_{\rm late}$ by using a simple relation, $v_{\eta} = (D_{p}-z_{\rm max})/t_{\rm late}$. For the sake of simplicity, we assume that the relaxation time to reach the terminal velocity is negligibly small. The estimated $\eta_{\rm late}$ values are shown in Fig.~\ref{fig:viscoelastic-results}(e,f). 

\subsection{Comparison with previous measurements}
The orders of magnitude of the obtained values $E_D = 10^6$--$10^7$~Pa, $\eta_D = 10^2$--$10^3$~Pa~s, and $\eta_{\rm late}=10^0$--$10^1$~Pa~s are consistent with previous measurements of the elasticity and viscosity of a dense suspension~(e.g. \cite{Maharjan:2018,Fall:2008}). While these previous studies solely measured elasticity or viscosity, we simultaneously measured them in a simple experiment. Moreover, the range of viscosity jump between $\eta_D$ and $\eta_{\rm late}$ is about two orders of magnitude (Fig.~\ref{fig:viscoelastic-results}(g,h)). Although the absolute value of the viscosity in a shear thickening suspensions varies significantly depending on the experimental conditions, the order of viscosity jump (2--3 orders of magnitude) is universal in a general DST suspension. The current experimental data are consistent with this viscosity-jump order. 

\subsection{Vibration effect}
\label{sec:vibration}
\begin{figure}
	\centering
	\includegraphics[width=0.9\linewidth]{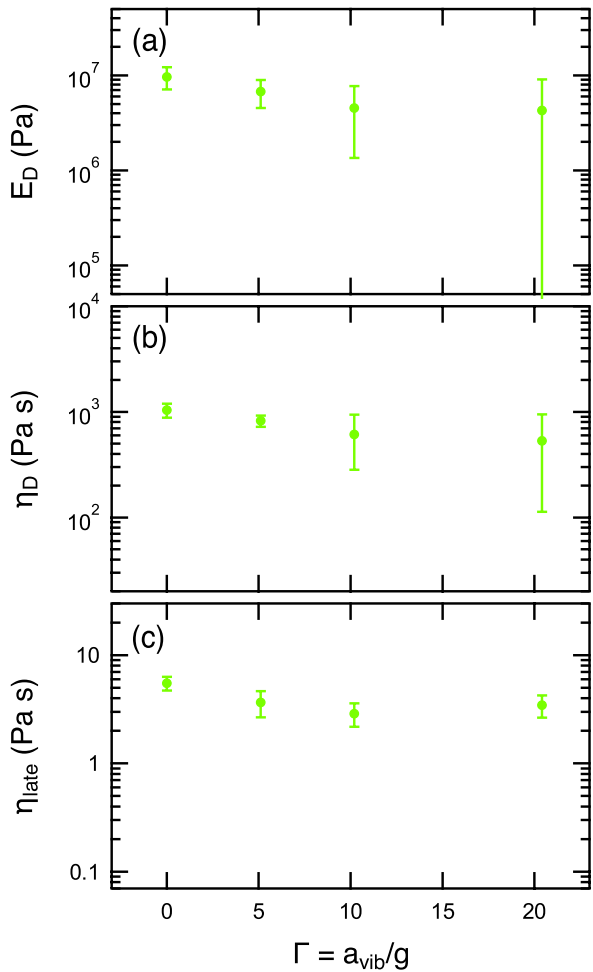}
	\caption[Vibration effect]{
	Vibration effect on impacted dense suspension. Relations among (a)~$E_{D}$, (b)~$\eta_{D}$, (c)~$\eta_{\rm late}$ and normalized maximum vibration acceleration $\Gamma=a_{\rm vib}/g$ are presented. Vibration frequency and target thickness are fixed at $120$~Hz and $20$~mm, respectively, in this experiment.}
	\label{fig:120Hz}
\end{figure}

To better understand the transient rheology of the impacted dense suspension, we perform additional experiments in which the vertical vibration is applied to the target dense suspension. Target layer is vertically vibrated by the vibrator (Fig.~\ref{fig:Experiment}). The main parameter in this experiment is the maximum vibration acceleration $a_{\rm vib}$. When $\Gamma~=~a_{\rm vib}/g$ exceeds around $10$, persistent holes and/or fingering of the dense suspension can be observed~\cite{Merkt:2004}.
We conduct the impact experiments under the effect of vibration in the range of $0~\leq~\Gamma~\leq~20$, and measure the viscoelastic properties using the method introduced thus far. 
Figure~\ref{fig:120Hz} shows the relations among the estimated $E_{D}$, $\eta_{D}$, $\eta_{\rm late}$, and $\Gamma$. From these plots, the elasticity and viscosity seem to decrease slightly with an increase in the maximum vibration acceleration. However, this tendency is not very clear. We do not observe a persistent hole or fingering before and immediately after the impact. The growth of fingers is gradually triggered by the impact when $\Gamma$ is sufficiently large.

\section{Discussion}
As discussed in previous studies~(e.g.~\cite{Waitukaitis:2012,Maharjan:2018}), we consider that a solid plug is developed by the jamming-front propagation induced by the impact. When the solid plug reaches the container bottom, the elastic stress transmits through the solid plug and results in a rebound. We assume that the stress transmission in the solid plug is much faster than the solidifying rate (jamming-front-propagation speed). The latter is usually comparable to the impact velocity ($\simeq 10^0$~m~s$^{-1}$ in this study). To check this comparability, we evaluate the ratio between $v_0$ and $H/t_{\rm stop}$ and find that the ratio is almost constant, $H/(v_0t_{\rm stop}) \simeq 4$, independent of the experimental conditions. On the other hand, the sound speed in the solid plug can be estimated as $\sqrt{E_{D}/\rho_f}\simeq 10^2$~m~s$^{-1}$, which is indeed much faster than $H/t_{\rm stop} = 10^0$~m~s$^{-1}$. Thus, the solid plug reaches the bottom within about $10^{-3}$~s. Then, the elastic spring is immediately developed within $10^{-5}$~s. Then, the rebound (elastic oscillation) occurs during $10^{-3}$~s ($t_{\rm res}-t_{\rm stop}$). This rebound timescale is determined by the effective elasticity of the solid plug. Therefore, there is no contradiction if we consider that the impact-induced solidification of the dense potato-starch suspension is caused by the dynamic jamming-front propagation. Once it reaches the bottom boundary, the elasticity causes the rebound of the projectile. From an instantaneous kinematic analysis with a linear model, the values of the effective elasticity and viscosity (above and below the DST) can be simultaneously estimated in this study. 

The idea of solid-plug formation is actually consistent with the behavior of $z_{\rm max} a_{\rm max}$ and $v_0$~(Fig.~\ref{fig:zmax-amax}). As mentioned in Sec.~\ref{sec:max-z-max-a}, the measured $z_{\rm max} a_{\rm max}$ is larger than the expected value. This large $z_{\rm max}a_{\rm max}$ cannot be explained by the added-mass effect. If we assume that $(m_{\rm i}+m_{\rm add})z_{\rm max}a_{\rm max}=m_{\rm i}v_{0}^2/2$ instead of Eq.~(\ref{eq:zav}), the relation becomes $z_{\rm max}a_{\rm max}=m_{\rm i}v_{0}^2/2(m_{\rm i}+m_{\rm add})$, indicating a smaller coefficient, $m_{\rm i}/2(m_{\rm i}+m_{\rm add})<1/2$. This tendency is inconsistent with the current experimental result~(Fig.\ref{fig:zmax-amax}(c)). 
By considering the delayed solidification effect~\cite{Maharjan:2018}, a relation $a_{\rm max}(z_{\rm max}-z_{\rm reach})\simeq v_0^2/2$ is obtained, where $z_{\rm reach}$ is the depth of the projectile at which the solid plug reaches the bottom. This form is qualitatively consistent with the curent experimental data. 

In Ref.~\cite{Maharjan:2018}, the effective modulus of the solid plug was measured by the constant-speed penetration of a solid cylinder into a dense corn starch suspension. Although the order of magnitude of the elasticity obtained in Ref.~\cite{Maharjan:2018} is close to the result obtained in this study, Maharjan et al. reported that the modulus increased as the fluid layer thickness $H$ increased~\cite{Maharjan:2018}. This tendency cannot be observed in our experiments. As seen in Fig.~\ref{fig:viscoelastic-results}(b), $E_D$ is independent of or a slightly decreasing function of $H$. This difference could be a result of the different loading conditions. In Ref.~\cite{Maharjan:2018}, the dense suspension was pushed at a constant speed. However, only the impulsive loading was added in this study. In the former, the solidified region is continuously compacted by constant-speed penetration, which can strengthen the solidified region even in a thick layer. In the latter, on the other hand, continuous compaction is impossible. Then, the weakening of the solidified region could be caused by the strong dissipation and attenuation of the jamming-front propagation in a dense suspension. This effect is enhanced in a thick layer. To prove the validity of this consideration, the loading-condition dependence of $E_D$ has to be carefully studied. This is an interesting future problem. 

In general, the mechanical properties of materials such as elasticity and viscosity should not depend on the size and/or boundary conditions. However, our experimental results clearly suggest that a bottom boundary is necessary to induce the rebound of the projectile, as advocated by \cite{Waitukaitis:2012,Maharjan:2018}. 
Actually, this constraint is not very peculiar because even a typical solid spring requires a fixed boundary condition to induce the oscillation (rebound). 
When $v_0<0.8$~m~s$^{-1}$ and H$=40$~mm, the rebound is not observed in our experiment. Since we focus on the viscoelastic characterization of the impacted dense suspension, this no-rebound regime is not analyzed in this study. A detailed study on the boundary and size effects on the effective elasticity is a challenging problem for future research. Moreover, we fixed the concentration (solid fraction) of the suspension. This could also affect the rebound properties. Thus, more systematic experiments are necessary to completely reveal the physics of the rebound on the impacted dense suspension.

Since the linear model used in this study (Eq.~(\ref{eq:VoigtModel})) is quite simple, there are some limitations. The most important quantity determining the mechanical properties is the timescale $T$. The orders of $E_D$ and $\eta_D$ are mainly determined by the order of $T$. As can be confirmed in Eqs.~(\ref{eq:E-ans}) and (\ref{eq:eta-ans}), $\ln (\epsilon)$ only affects the factors of $E_D$ and $\eta_D$. In the current analysis, timescale $T$ is determined by $t_{\rm stop}$ and $t_{\rm res}$. To obtain the fitting curve shown in Fig.~\ref{fig:Kinematic-data-sets}(b), the oscillatory phase and initial condition are set to the fitting parameters, while $E_D$ and $\eta_D$ are fixed to the estimated value. Then, the data behavior within $(t_{\rm stop},t_{\rm res})$ can be fitted (Fig.~\ref{fig:Kinematic-data-sets}(b)). However, a full waveform cannot be reproduced by the model of Eq.~(\ref{eq:VoigtModel}). In particular, on a longer timescale, the effects of gravity, capillary force, and viscosity relaxation cannot be neglected. As a consequence, relatively slow motion (whose timescale is much longer than $T$) and an equilibration-level offset ($z\simeq 2$~mm), which are inconsistent with the model, are observed after $t_{\rm res}$. In the very early stage, on the other hand, the elastic response is not relevant because the solid plug does not reach the bottom. In this range, the added-mass effect dominates the deceleration. Although this deceleration is effectively inclded in the viscosity in the model, its effect is limited. Owing to these limitations, the linear viscoelastic model is only applicable to a short duration of around $(t_{\rm stop},t_{\rm res})$. However, this duration is the most important part in analyzing the elastic response. Therefore, we focus on this range and use the simple model.  

As mentioned previously (in Sec.~\ref{sec:introduction}), the term ``elasticity'' usually refers to the energy storage. However, in an impacted dense suspension, the energy is not stored in the solid plug~\cite{Maharjan:2018}. Thus, $E_D$ merely indicates the pseudo (effective) elasticity, which characterizes the solid-like behavior of the impacted dense suspension that causes a dissipative rebound.

If we can weaken (or relax) the solid plug structure, the rebound can be suppressed. A simple idea to weaken the solidification is applying a perturbation. This is why we carried out the impact experiment under the effect of a target vibration. However, as shown in Fig.~\ref{fig:120Hz}, the effect of vibration to weaken the solid plug is quite limited. The solid plug structure is rather stable. The viscoelastic features of the impacted dense suspension are not significantly affected by the mechanical vibration. However, since we fixed the vibration frequency to $f=120$~Hz, the frequency dependence is not investigated in this study. In addition, the direction of vibration can affect the viscoelasticity~\cite{Lin:2016}. More systematic measurement is needed to conclude the effects of vibration.

\section{Conclusion}
We performed a simple impact experiment with a steel sphere impinging on a dense potato-starch suspension. When the thickness of the target-suspension layer was shallow enough and the impact velocity was high enough, a rebound of the projectile was observed. To characterize the effective viscoelasticity of the target suspension, we employed a simple linear viscoelastic (Voigt) model. Using the model and the kinematic data of the projectile motion, we estimated the effective elasticity $E_D$ and viscosity $\eta_D$ above the DST transition. By measuring the slow sinking timescale, we additionally estimated the viscosity $\eta_{\rm late}$ below the DST transition. The obtained values were consistent with those in previous studies in which the viscosity and elasticity were measured separately. To explain the rebound behavior, the dynamic jamming front must reach the bottom boundary. This implies that the existence of a bottom boundary is indispensable to make an elastic response that causes the rebound. In addition, we examined the effect of mechanical vibration on the effective elasticity and viscosity. As a result, the measured elasticity and viscosity are almost independent of the vibration strength. This means that the impact-induced solidification is stable against mechanical perturbation.

\section*{Acknowledgement}
We thank Ryohei Seto and Scott Waitukaitis for valuable discussions and for introducing useful references. This work was supported by JSPS KAKENHI grant no.~18H03679. 

\bibliography{dst}

\end{document}